\documentclass[english]{article}

\usepackage[T1]{fontenc}
\usepackage[latin9]{inputenc}
\usepackage{amssymb}
\usepackage{babel}

\usepackage[pdftex]{color,graphicx}

\begin{document}

\title{Firewall or smooth horizon?}

\author{Amos Ori\footnote{\tt amos@physics.technion.ac.il}
\\
\\ \small{Department of Physics} \\ \small{Technion-Israel Institute of Technology} \\ \small{Haifa 3200, Israel}}

\date{\today}

 \maketitle

\begin{abstract}
Recently, Almheiri, Marolf, Polchinski, and Sully found that for a
sufficiently old black hole (BH), the set of assumptions known as
the \emph{complementarity postulates} appears to be inconsistent with
the assumption of local regularity at the horizon. They concluded
that the horizon of an old BH is likely to be the locus of local irregularity,
a {}``firewall''. Here I point out that if one adopts a different
assumption, namely that semiclassical physics holds throughout its
anticipated domain of validity, then no inconsistency seems to arise,
and the horizon retains its regularity. In this alternative view-point,
the vast portion of the original BH information remains trapped inside
the BH throughout the semiclassical domain of evaporation, and possibly leaks
out later on. This appears to be an inevitable outcome of semiclassical
gravity. 
\end{abstract}

\section{Introduction}

Almheiri, Marolf, Polchinski, and Sully \cite{AMPS} recently analyzed
the evaporation of black holes (BHs) via Hawking radiation \cite{Hawking},
from the view-point of information theory. They concluded that, if
one accepts a set of postulates concerning evaporating BHs (the so-called
\emph{complementarity postulates}), including the assumption that
local physics is regular at the horizon, one is led to a contradiction.
Based on this contradiction, AMPS proposed that for a sufficiently
old evaporating BH, the horizon's local regularity predicted by classical
(and semiclassical) theory is actually replaced by a pathological
local behavior at the horizon---a \emph{firewall}. This conclusion
was subsequently confirmed by several authors \cite{agree(susskind),Giveon}, 
though several other authors disagree 
\cite{Mathur,Weinberg(disagree),Banks(disagree)}. 
(See also \cite{Bousso(disagree),Chowdhury}.) 

It must be noted that no mechanism was proposed in \cite{AMPS} for
explaining how the irregularity actually develops: It is merely the
alleged contradiction found in the set of {}``complementarity postulates''
(combined with the assumption of regular horizon) that led AMPS to
the firewall proposal. 

One should also bear in mind that the analysis and discussion in \cite{AMPS,agree(susskind)}
rely on a certain assumption, which is so commonly used that the authors
do not even bother to spell it out in their set of postulates: 
Namely, that most of the BH information
is carried out with the Hawking radiation, already in the semiclassical
phase of evaporation. We may refer to this additional assumption as
the {}``zeroth complementarity postulate.'' %
\footnote{One may contend that this additional assumption is implicitly contained
in the first complementary postulate. (However, this is not necessarily
the most natural interpretation of that postulate, as usually formulated.) %
} 

The main goal of this manuscript is to discuss this problem from a
different view-point. Instead of the complementarity postulates, I
will simply assume that the semiclassical theory of gravity (augmented
by the equivalence principle) holds throughout its anticipated domain
of validity---that is, as long as curvature is small compared to Planckian
value. I will argue that this assumption naturally leads to a different
scenario, in which most of the information is actually stored inside
the shrinking BH throughout the semiclassical stage of evaporation,
and may possibly be released later on (after the BH approaches a small
mass, of order the Planck mass $m_{p}$). The contradiction pointed
out by AMPS does not arise in this scenario, hence there is no reason
to assume a firewall: Instead, the horizon is regular throughout the
semiclassical domain---that is, as long as the BH is macroscopic ($m\gg m_{p}$);
and seemingly no inconsistency is encountered. 

Throughout the manuscript I will consider an uncharged, non-spinning,
spherical BH, and use the Planck units $c=G=\hbar=1$.

A key ingredient in this discussion is Bekenstein's BH entropy, $S=4\pi m^{2}$
(in Planck units). It may be tempting to interpret this S as characterizing
the number of different micro-states associated with a macroscopic
BH of given mass $m$ --- or, stated in other words, as a measure
of the overall {}``amount of information'' carried by the BH. We
shall later demonstrate, however, that this interpretation becomes
very problematic when applied to an evaporating BH with remaining
mass $m$ (as long as the validity of semiclassical physics is assumed).
This is in fact a fairly well-known observation, though seemingly
it is often overlooked in current literature. 

Over the last two decades it became widely accepted that (unlike Hawking's
original claim) information is actually conserved in the process of
BH evaporation---and more specifically, that an initial pure state
remains pure. Various approaches have been proposed as for how the
information is encoded in the BH, and at what form (and at what timing
and rate) is it released to the external world. 

Consider now an evaporating BH with initial mass $M_{0}$. According
to semiclassical theory the remaining BH mass $m$ evolves according
to \[
m(v)=c(t_{0}-v)^{1/3}\]
where $c$ is a certain constant, $v$ is the Eddington outgoing coordinate
(parametrizing the BH horizon, and set to zero at the moment of collapse),
and $t_{0}$ is a constant characterizing the BH's life-time, from
collapse to full (semiclassical) evaporation: $t_{0}=(M_{0}/c)^{3}$.
This law presumably applies as long as the remaining BH mass is macroscopic,
that is, $m(v)\gg m_{p}$. 

The various proposals as for how information is stored in the evaporating
BH, and at what time (and rate) is it released out, may be crudely
divided into two categories:
\begin{description}
\item [{(A)}] The BH entropy $S(v)=4\pi m(v)^{2}$ properly represents the (steadily
shrinking) information capacity of the evaporating BH. Accordingly,
assuming that no information loss occurs, we must assert that while
the BH evaporates, the decrease in $S(v)$ is directly translated
into increase (by the same amount) of the information contained in
the Hawking-radiation field (this information is encoded in inter-correlations
between emitted particles, as well as correlations with the BH near-horizon
and/or internal state). Stated in other words, the amount of information
contained in the Hawking radiation up to an {}``advanced moment''
$u$ is 
\begin{equation}
I(u)\equiv S(M_{0})-S(m_{b}(u))=4\pi \left[M_{0}^{2}-m_{b}(u)^{2}\right]
=4\pi c^{2}\left[t_{0}^{1/3}-(t_{0}-u)^{1/3}\right]^{2},\label{eq:Info}\end{equation}
where $u$ denotes the ingoing Eddington coordinate (given by $u=v-2r_{*}$),
and $m_{b}(u)$ is the Bondi mass (namely, the remaining mass, as
observed by a far observer as a function of $u$). %
\footnote{A simple intuitive way to conceive this quantitative notion of {}``information''
(and its relation to $I$) is to count the number of binary bits required
for encoding all possible micro-states of the system in consideration.
This is of course closely related to the notion of Entropy, which
is the log of the number of micro-states. Thus, $I$ (as defined in
Eq. (\ref{eq:Info})) may be interpreted here as the number of required
bits, multiplied by $\ln2$. %
} Thus, $I$ vanishes at $u=0$ (the advanced moment of collapse),
and grows monotonically with $u$ until it gets the maximal value
$I_{max}=4\pi M_{0}^{2}$ at the end of evaporation, $u=t_{0}$. (Note
that when the remaining Bondi mass becomes small, $m_{b}(u)\ll M_{0}$,
$I$ is approximately $I_{max}$ --- even though $m_{b}$ is still
macroscopic.) We may refer to this scenario as \emph{prompt information
release}. 
\item [{(B)}] The information encoded in an evaporating BH with remaining
mass $m$ may be much larger than $S(m)$, and in fact it is by no
means bounded by the current value of $m$. According to this view,
the Hawking radiation may encode (throughout the semiclassical phase
of evolution) very little information about the initial state of the
system. Most of the information thus remains trapped inside the shrinking
BH. This scenario, to which I will refer as \emph{confined information},
comes in several variants, differing by the final fate of the information
(and of the internal BH geometry): (B1) The information may be trapped
for a long time in a small-mass (say, $m\sim m_{p}$) remnant of finite
life-time, and be released during a long period of time, after the
completion of (the semiclassical stage of) the BH evaporation. I will
refer to this variant as \emph{delayed information release}; or (B2)
it may be trapped forever in a stable small-mass ($m\sim m_{p}$)
remnant, or (B3) perhaps it may migrate to a {}``baby universe''
(created when the internal part of the small-$m$ BH pinches off our
parent universe). 
\end{description}
Over the last two decades, the first view-point (A) became the most
dominant one. An important boost emerged from analyses by Bekenstein
\cite{Bekenstein} and Page \cite{Page}, showing that, despite of
the approximate thermal character of the Hawking radiation, its potential
information capacity is large enough to carry the entire BH information.
Later this idea was further boosted due to the analysis by Hayden
and PresKill \cite{Hayden}. One should bear in mind, however, that
none of these works provided any concrete indication that information
is actually carried out by the emitted radiation (namely, that the
state of the radiation field is actually correlated with the BH's
initial micro-state). 

The analysis in Ref. \cite{AMPS}, and similarly the discussion in
\cite{agree(susskind)}, rely on this assumption of prompt information
release (scenario A) --- and so is their conclusion concerning the
existence of firewalls. 

On the contrary, the discussion in this manuscript will instead rely
on one key assumption, to which I will refer as the \emph{semiclassicality
postulate}: 
\begin{itemize}
\item The semiclassical theory of gravity holds as long as curvature is
much smaller than Planckian. (That is, along the horizon semiclassical
theory should apply as long as $m\gg m_{p}$, and inside the BH it
should apply in all regions which are not too close to the $r=0$
singularity.) 
\end{itemize}
Stated in other words, we postulate that no mysterious unexpected
phenomena are to be expected to occur at the macroscopic level. In
particular this postulate would forbid non-local macroscopic phenomena---and
obviously non-causal ones. This is surely a rather conservative assumption. 

In particular, it will be assumed that in the semiclassical domain
there is a well-defined notion of (classical) geometry, with a well-defined
causal structure, and this geometry satisfies the Einstein equations
with the usual semiclassical source term $\hat{T}_{\alpha\beta}$,
the renormalized stress-Energy tensor. Below I will argue that once
this postulate is adopted, one is naturally led to the second view-point
(B). The consequence is that no mysterious firewall is needed, and
the horizon may still be viewed as a regular place---as suggested
by the equivalence principle (and by semiclassical gravity). 

I should emphasize that this picture (B) is by no means new: Some
variants of it were proposed several decades ago, for example by Banks
and collaborators \cite{horn}. It was also proposed by Parentani
and Piran \cite{Piran}. More recently this scenario was re-considered
by Hossenfelder and Smolin \cite{Smolin}. Yet, for some reasons (probably
due to some objections which I'll mention below), this view-point
never gained much popularity. My feeling, however, is that the very
recent dramatic developments---the recognition that the complementarity
postulates are not contradiction-free (and the resultant firewall
argument) call for a re-consideration of this alternative view-point
(B). I find this view-point quite compelling.

\section{Semiclassical spacetime of evaporating black hole }

Consider the gravitational collapse of an uncharged, spherically-symmetric,
compact object with large initial mass $M_{0}$. The resulting BH
then evaporates during a very long time, $\propto M_{0}^{3}$. Figure
1 displays the corresponding spacetime diagram. 

\begin{figure}[ht]
\begin{center}
\includegraphics[scale=.4]{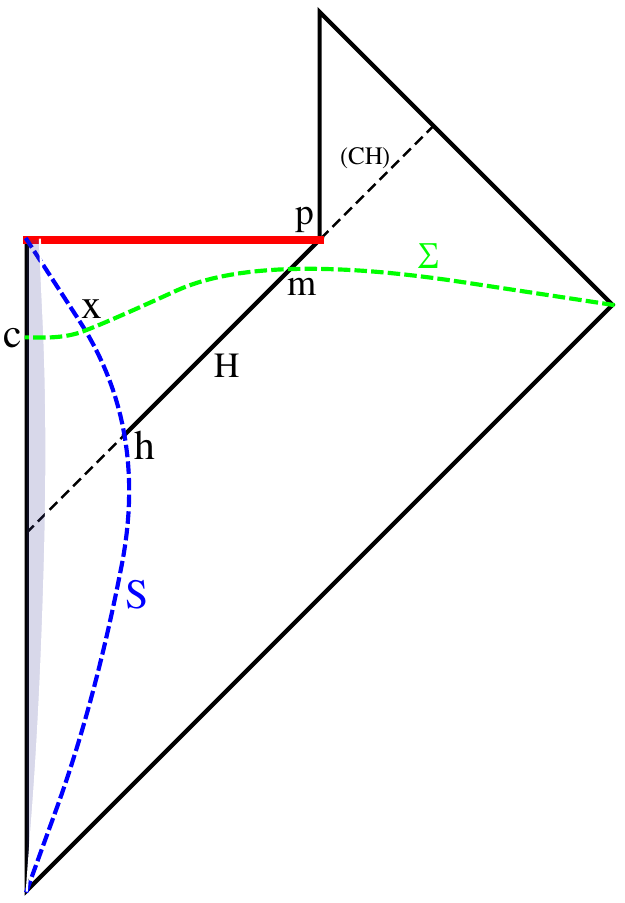}
\caption{\label{simple}
Spacetime diagram of an (uncharged) evaporating spherical BH. The
horizontal red line denotes the $r=0$ singularity, the dashed blue
line S is the worldline of the collapsing thin shell, and the bold
diagonal black line marked by H is the event horizon. The point {}``p'',
the intersection of the horizon and the $r=0$ singularity, is the
{}``end of evaporation'' point, which is actually a (naked) singularity
of the semiclassical spacetime. The diagonal dashed line marked by
{}``(CH)'' (which extends the horizon in the upper-right direction)
is in fact a Cauchy horizon of the semiclassical spacetime, beyond
which the extension of geometry is not really predictable by the semiclassical
theory. The green dashed line denotes the spacelike hypersurface $\Sigma$
(see text). The points marked by c, x, and m respectively denote the
intersection points of $\Sigma$ with the regular center, with the
collapsing shell, and with the horizon (at a {}``moment'' $v$ where
the remaining mass is $m$, much smaller than the initial mass $M_{0}$).}
\end{center}
\end{figure}

To simplify the discussion, we shall consider the case of a thin-shell
collapse; however, our main conclusions will as well apply to the
more general and more realistic case of e.g. a collapsing star. 

Upon evaporation, the BH mass $m(v)$ monotonically shrinks from $M_{0}$
toward smaller values. Let us discuss the situation at a fairly late
advance time $v=v_{m}$, where most of the mass has already been evaporated,
yet the remaining BH mass is still macroscopic: $m_{p}\ll m\ll M_{0}$.
To this end we choose a typical spacelike hypersurface $\Sigma$ which
intersects the horizon at $v=v_{m}$ (the point denoted {}``m''
in Fig. 1). One can easily construct such a spherically-symmetric
hypersurface $\Sigma$ which is asymptotically-flat, smooth everywhere
(except, technically speaking, at the thin shell), and with (intrinsic
as well as extrinsic) curvature which is nowhere larger than $\sim1/m^{2}$.
Such a hypersurface $\Sigma$ is depicted in Fig. 2, by plotting the
corresponding area coordinate $r$ as a function of proper length
$L$ (in the radial direction). An embedding diagram of the hypersurface
$\Sigma$ is shown in Fig. 3. 

\begin{figure}[ht]
\begin{center}
\includegraphics[scale=.5]{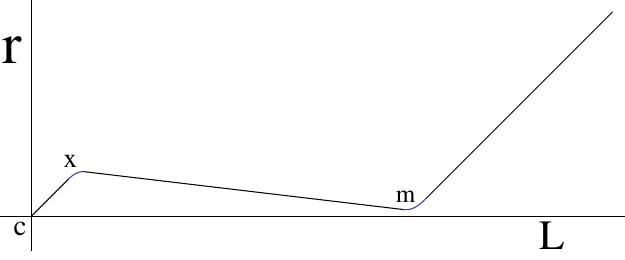}
\caption{\label{simple}
The geometry of the spacelike hypersurface $\Sigma$, represented
her by the corresponding function $r(L)$, where $r$ is the area
coordinate and $L$ is the proper length along the hypersurface $\Sigma$
(in the radial direction). The points c, x, and m (see caption of
Fig. 1) are marked.}
\end{center}
\end{figure}

\begin{figure}[ht]
\begin{center}
\includegraphics[scale=.5]{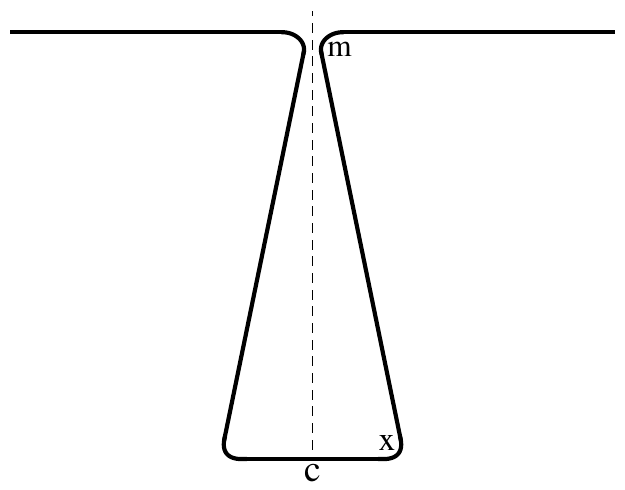}
\caption{\label{simple}
Embedding diagram for the hypersurface $\Sigma$. Note the narrow
throat (at the horizon-crossing point {}``m'') and the large internal
volume. 
}
\end{center}
\end{figure}

{[}Here is a sketch of a simple construction of such a hypersurface
$\Sigma$: (i) Inside the shell, take it to be a hypersurface $t_{m}=const$,
where $t_{m}$ denotes the Minkowski time inside the shell---that
is, take $r(L)=L$. (ii) Along the long section of $\Sigma$ from
point {}``x'' to point {}``m'', take $r=\gamma m_{e}(L)$, where
$0<\gamma<2$ is a constant that may take any value which is not too
close to zero or $2$ (we may take for example $\gamma\sim1$), and
$m_{e}$ denotes the ($L$-dependent) effective BH mass. (Due to the
BH evaporation $m_{e}$ varies monotonically from $M_{0}$ at point
{}``x'' to $m$ at point {}``m''.) (iii) In the neighborhood of
the horizon-crossing point {}``m'', take the curve $r(L)$ to be
a radial spacelike geodesic (in the Schwarzschild geometry with mass-parameter
$m$). It is described by the solution of the equation \begin{equation}
\frac{dr}{dL}=\left[\frac{2}{\gamma}-\frac{2m}{r}\right]^{1/2}.\label{eq:r(L)}\end{equation}
In such a solution $r(L)$ grows monotonically from $r=\gamma m$
at a certain $L$ (inside the BH) up to the asymptotic external region
$r\gg2m$ where it takes a constant-slope asymptotic form, $r\cong(2/\gamma)^{1/2}L+const$.
(iv) Beyond a certain $r\gg2m$, pick a linear function $r=L+const$.
It is not difficult to smoothly glue these four different functions
$r(L)$ to each other, by slightly deforming them near their respective
boundaries.{]}

Notice the fairly unusual geometric shape of $\Sigma$: It has a narrow
throat of radius $r\approx2m$ (located at point {}``m'' in Figs.
2,3). However, {}``inside'', to the left of the throat, there is
a much larger, elongated, {}``balloon'' of typical width $\sim M_{0}$
(and typical length $ $$\sim M_{0}^{3}$). It thus admits a huge
3-volume. It resembles the {}``horned'' configurations previously
considered by Banks and collaborators \cite{horn} in some respects,
although there are several important differences. 

We point out that no exact solutions are known for the semiclassical
field equations describing the spacetime of a 4D evaporating BH. Furthermore,
even the form of $\hat{T}_{\alpha\beta}$ is not known explicitly
in 4D (for a prescribed time-dependent metric). The picture portrayed
if Figs. 1-3 relies on several inputs: (i) Qualitative considerations
taking into account some known key features of $\hat{T}_{\alpha\beta}$
in 4D (and using the adiabatic approximation, applicable to macroscopic
BHs, for constructing the semiclassical geometry); (ii) Detailed numerical
analysis of spherical collapse and evaporation in 4D, carried out
by Parentani and Piran \cite{Piran} (who \emph{assumed} a specific
form of $\hat{T}_{\alpha\beta}$); and (iii) Numerical \cite{Dori}
as well as approximate-analytical \cite{Aprox} investigation of the
analogous problem in 2D gravity, based on the CGHS model \cite{CGHS}.
(Note that in 2D, $\hat{T}_{\alpha\beta}$ is known explicitly for
any prescribed metric \cite{CGHS}.) These three pieces of information
all lead to the same qualitative structure of spacetime --- a large-scale
region located inside the BH, beyond a narrow throat --- as depicted
in Fig. 3. %
\footnote{Yet, there are certain differences between the 2D and 4D semiclassical
spacetimes. Most notably, in 4D $\Sigma$ admits a regular center,
i.e. the point c. Instead, in 2D $\Sigma$ extends to the left up
to a second spacelike infinity (disconnected from the one at the right
edge). %
}

\subsection{The hard-disc thought experiment}

How much information can be stored in the remaining BH when it is
so old that its remaining mass $m$ is $\ll M_{0}$? It might be tempting
to postulate that this amount is of order $S(m)=4\pi m^{2}$, and indeed,
such statements often appear in the literature. However, at least
according to our conservative semiclassicality postulate, this cannot
be the case! In fact, a spacelike hypersurface like $\Sigma$ admits
a huge 3-volume---regardless of how small $m$ is, and it may therefore
harbor a huge amount of information. 

To verify the last point, let us carry the following thought experiment:
Recall that our model assumes a collapsing thin shell, with (approximately)
Minkowski interior. Suppose we place a $10$-Terabyte hard-disc %
\footnote{What I actually have in mind, when saying {}``hard disc', is a disc
of hard metal, on which the information is scratched, encoded in a
certain binary code (so in a sense it is reminiscent of an optical
{}``compact disc'', or a gramophone disc). So one can actually think
of this {}``hard disc'' as a static device, which does not need
to consume energy --- which simplifies the considerations below.%
} inside the shell---say, in its central region, the region marked
by gray in Fig. 1 (near its left boundary). As long as this hard-disc
is still intact, we are assured that the BH information capacity is
at least $10$-Terabyte (multiplied by $\ln2$). For, the hard
disc can be in $2^{(10^{13})}$ different states, all with same mass. 

(Of course, the BH's information storage is much larger than this:
Any prescribed state of the system of $10^{13}$ {}``mesoscopic''
bits can be represented by many different micro-states. However,
for the sake of the present discussion, the derivation of a lower
bound on the amount of information capacity, it will be sufficient
to recall that the system now has at least $10$-Terabyte information
capacity.) 

The spacelike hypersurface $\Sigma$ intersects the center of symmetry
at the point $c$ in Fig. 1. The hard disc, with its $10^{13}$-bits
information capacity, is thus registered at the {}``moment'' $\Sigma$.
The disc is still intact (and hence the stored information is still
safe), because virtually no tidal forces are present in the (almost)
flat interior of the shell. Thus, even when the evaporating BH remains
with arbitrarily small (though still macroscopic) mass $m$, it still
enjoys this information capacity of the hard disc. 

Obviously, we can actually store inside the hollow collapsing shell
\emph{many} hard discs of this type. We are only restricted by the
overall mass of the discs, which will be limited by the initial BH's
mass $M_{0}$. %
\footnote{To be more pragmatic, we demand that the total mass of the hard discs
will be bounded by $\alpha M_{0}$, where $\alpha$ is a certain fixed
parameter, $\ll1$. 

(One can also show that for fixed $\alpha$, the {}``hydrostatic
pressure'' emerging from the self-gravity of the system of hard-discs,
if distributed homogeneously inside the shell, will decrease with
$M_{0}$ like $M_{0}^{-2}$.)%
} One thus finds that the amount of information that may be stored
in hard-discs and survive up to $\Sigma$ is bounded below by \begin{equation}
I_{disc}=\beta M_{0}\label{eq:Idisc}\end{equation}
where $\beta$ is some constant. 

We therefore conclude that \emph{the information stored in an evaporating
BH is by no means restricted by its remaining mass} $m$! It may only
be restricted by its \emph{initial} mass $M_{0}$. 

Of course, we do not really need a hard disc to store the information
concerning the initial state. We can replace the thin shell and the
hard disc altogether by e.g. a collapsing star. The matter composing
the star should carry (almost) all the information concerning the
system's initial state, encoded in its (time-evolving) micro-state
--- at least until it gets very close to the singularity. 
(Later I'll briefly discuss the fate of information when the hard disc hits the
$r=0$ singularity.) I chose to establish the toy model on a {}``hard-disc''
device because it has several simplifying ingredients: The information
is stored in a robust and durable manner, scratched on a hard metal
disc. It is well-localized, and hence well protected from damage as
long as the disc is still inside the hollow shell. Also the information
may be accessed in a direct, non-intricate manner (as opposed to e.g.
subtle correlations between photons or atoms), so no fine subtleties
are involved in an attempt to access the stored information. Furthermore,
since the scratches are in fact \emph{macroscopic}, avoiding the above
conclusion would require violations of the principles of locality
and causality \emph{already at the macroscopic level} (see discussion
below). %
\footnote{$ $The hard-disc toy model yielded an information lower bound $I_{disc}$
proportional to $M_{0}$. A more realistic physical system (e.g. a
collapsing star) should probably yield a significantly larger information
capacity. For example, considering a gas of photons enclosed in the
collapsing shell, one obtains a lower bound $\propto M_{0}^{3/2}$. %
} 

We must emphasize that this simple observation (namely that the information
carried inside the BH is not sensitive to the remaining mass $m$)
is not new. It was already pointed out, for example, by Preskill \cite{Preskill}
and several others, several decades ago. In fact it was this observation
of the non-shrinking internal information capacity which led Susskind
and collaborators \cite{complementarity} to introduce the logically-subtle
concept of {}``complementarity'': If for some reason one is willing
to insist that the information concerning the BH's initial state is
released out of the BH along with its entropy $S(m)$, then, by the
time the remaining mass $m$ is $\ll M_{0}$, one faces the difficulty
of holding (on $\Sigma$) \emph{two copies} of the information: One
deep inside the BH (e.g. our hard disc), and the other one in the
emitted radiation field. This would conflict with the quantum-mechanical
{}``no-cloning'' principle. The complementarity concept was aimed
to resolve this conflict. But of course there is a logically simpler
way out of this conflict: There is no need to assume in the first
place that the bulk of BH information is emitted together with the
BH entropy. 

What will happen to the information stored in the hard disc (or alternatively
in the collapsing star) at later stages, when the disc heads towards
the $r=0$ singularity? At some stage, the collapsing shell will presumably
crash on the disc, and the latter will ultimately be destroyed (as
a mechanical device) --- due to this clash, and also due to the growing
tidal forces. This does not mean that information is really destroyed
at this stage: rather, it should merely take a more subtle and intricate
form. This assertion, which is based on standard rules of local physics
(augmented by the equivalence principle) should presumably hold as
long as curvature is not too large. But it is less clear what will
happen to the matter debris afterward, e.g. when curvature grows to
Planckian values and even beyond (formally the curvature will diverge
at some stage, when the debris' wolrdlines approach the $r=0$ singularity).
One might suppose that perhaps information will truly be destroyed
at this stage. But this is still not necessarily the case: It has
been suggested by various authors that Quantum Gravity will somehow
cure the classical singularity, hence avoiding true information destruction.
An especially interesting scenario demonstrating this idea was proposed
several years ago by Ashtekar, Taveras and Varadarajan \cite{ATV},
based on quantization of the CGHS \cite{CGHS} semiclassical model. 

Is there any reasonable way to evade the above conclusion, the $m$-independence
of the information capacity? As Preskill \cite{Preskill} noted, this
would require the existence of a strange {}``bleaching'' mechanism
that {}``strips away (nearly) all information about the collapsing
body as the body falls through the apparent horizon (and long before
the body reaches the singularity)''. Our collapsing-shell model makes
it clear, however, that this {}``bleaching'' must take place way
\emph{before} the disc arrives at the apparent horizon (which coincides
in this region with the shell's hypersurface; see Fig. 1). Such a
mysterious {}``bleaching'' phenomenon conflicts with the equivalence
principle, and represents a macroscopic violation of locality \cite{Preskill}.
Moreover, one can deform $\Sigma$ slightly toward the past at its
left side, such that the point c will no longer be in the causal future
of h. In other words, the hard disc now intersects $\Sigma$ before
it can get any causal signal from any point on the apparent horizon.
Therefore, the {}``bleaching'' phenomenon would also require a violation
of causality, already at the macroscopic level.

\subsection{Conclusion: Information content of the shrinking BH}

Summarizing, the discussion in the previous subsection led to a simple
conclusion (based on the semiclassicality postulate):
\begin{itemize}
\item The information enclosed inside an evaporating BH is not restricted
in any way by the current value of $m$ (the remaining BH mass). This
seems to be an inevitable result---unless one is willing to postulate
the existence of a mysterious {}``bleaching'' mechanism (which,
as noted above, involves a violation of causality at the macroscopic
level, well within the semiclassical domain; and ultimately it would
violate our semiclassicality postulate). 
\end{itemize}
Once this conclusion is accepted, it has a direct logical consequence:
\begin{itemize}
\item There is no need to assume that a significant amount of information
is emitted through Hawking radiation. Instead, we may contend (as
Hawking did originally) that during the semiclassical stage of evaporation,
the radiation field carries very little information---much smaller
than the original capacity $4\pi M_{0}^{2}$. 
\end{itemize}
Thus, the semiclassicality postulate naturally leads to option B in
the discussion: Namely, the original BH information stays trapped
inside the BH throughout the semiclassical phase of evaporation.

This view-point B has obvious advantages: It is no longer necessary
to introduce the logically-complex concept of {}``complementarity''.
Furthermore, the apparent contradiction which led AMPS to introduce
the firewall concept is now avoided.

\section{What is the final fate of information?}

Assuming that the vast of BH information is not emitted during the
(semiclassical phase of) Hawking evaporation, there may be several
options concerning its final fate: 
\begin{description}
\item [{(i)}] Information is stored in a small-mass ($m\sim m_{p}$) remnant
of finite life-time. 
\item [{(ii)}] Information is stored in a small-mass stable remnant; 
\item [{(iii)}] Information may leaks to a {}``baby universe''.
\end{description}
Each of these options has its own advantages and difficulties. An
excellent discussion of these scenarios may be found in Ref. \cite{Preskill}. 

I find option (i) to be the most compelling one. Information is stored
for a long time in a temporary $m\sim m_{p}$ remnant. The latter's
(time-evolving) internal geometry resembles that of $\Sigma$ (depicted
in Figs. 2,3): It has a rather narrow throat (presumably of order the
Planck size), but to the left of it there is a large, macroscopic,
internal {}``balloon'', inside which the information is stored.
This internal balloon presumably shrinks with time slowly, {}``pouring
out'' its information content to the external universe, through massless
particles of extremely low frequency (typically $\omega\sim M_{0}^{-2}$
in Planck units, or possibly even smaller). The system's end-state
will then be just radiation field. Note that in this scenario the
radiation field is composed of two components: The standard Hawking
radiation (emitted in the semiclassical phase of evaporation), and
the subsequent, {}``post-evaporation'' component. Overall, the radiation
field will presumably be in pure state --- just like in the more popular
scenario of {}``prompt information release'' (scenario A in the
Introduction). 

It is sometimes argued \cite{Giddings}, based on the approach of
effective field theory, that if small-mass remnant configurations
are allowed, then spontaneous pair-production of such remnants must
occur with unbounded rate. Several authors pointed out, however, that
the applicability of effective field theory to this problem is in
question, due to the large-scale, macroscopic nature of the remnant's
internal geometry \cite{horn,Smolin,PreskillA}. 

In this regards one should also bear in mind that the formation of
{}``balloon''-like configurations like $\Sigma$ is in fact \emph{guaranteed}
by the semiclassical laws of evolution, and these laws also guarantee
that the throat will continue to shrink, presumably up to Planck
scale (at least). So small-mass balloon-like configurations admitting narrow (Planck-scale)
throat and large macroscopic interior are not only legitimate elements
in superspace, but also real physical states that must form in black-hole
evaporation (provided that we accept semiclassical gravity throughout
the domain of sub-Planckian curvature). The only question is the final
fate of these configurations: Stable remnant, or a temporary one,
or perhaps a baby universe.

\section{Discussion}

Our discussion throughout this manuscript was based on one key assumption,
the {}``semiclassicality postulate'': Namely, that semiclassical
physics (along with the equivalence principle) applies throughout
its anticipated domain of validity. In particular, no causality violation
is to be anticipated (especially at the macroscopic level) in the
semiclassical domain. Using this assumption, and focusing attention
on the state of matter inside the BH, we have arrived at the following
conclusion: The information content of an evaporating BH with a (macroscopic)
remaining mass $m$ cannot be bounded by the value of $m$ in any
way (though it may be bounded by the BH's initial mass $M_{0}$).
In turn, this observation naturally leads us to option B in the Introduction:
Namely, that the Hawking radiation does not carry out a significant
amount of information (throughout the semiclassical domain of evaporation). 

As a consequence, no firewalls are expected to develop at the horizons
of old evaporating BHs: The contradiction that led AMPS to propose
the firewall scenario was based on a set of assumptions, including
the presumption that the BH information is efficiently carried out
by the Hawking radiation, along with the BH mass. This presumption is now avoided. Instead
we assume that the semiclassical theory of gravity holds (which in
particular implies regularity of the horizon), and no inconsistency
seems to arise. 

In the scenario which emerges here, most of the original BH information
is stored in the BH interior (instead of being Hawking-radiated out).
This view is also motivated by the geometric properties of a typical
hypersurface $\Sigma$ that intersects the horizon at a moment $v$
of small remaining mass $m$. The 3-geometry of $\Sigma$ is characterized
by a narrow throat ($\sim m$) but a large internal volume (essentially
independent of $m$), with typical radius $\sim M_{0}$ and typical
length $\sim M_{0}^{3}$, so the volume is huge, of order $(M_{0}/m_{p})^{2}M_{0}^{3}$. 

There may be several options concerning the fate of this information
after the BH completes its evaporation \cite{Preskill} (see previous
section). A compelling option is that of a finite-lifetime remnant
(scenario B1 in the Introduction): The information is temporarily
stored in a finite-lifetime, small-mass ($m_{rem}\sim m_{p}$) remnant,
and slowly released through massless particles of very small frequency,
of order $M_{0}^{-2}$. The end state should thus be a radiation field
in pure state (assuming that the system's initial state was pure). 

The small-mass remnant scenario was discussed by several authors previously,
but it never gained popularity. This is probably due to several reasons:
(i) For {}``esthetic'' reasons, it may be tempting to assume that
the information capacity of an evaporating BH with remaining mass
$m$ is just $S(m)\approx 4\pi  m^{2}$ --- which (for small $m$) is too
small to store all the initial information; (ii) As was already mentioned
in the previous section, it is often argued \cite{Giddings} that
if small-mass remnant configurations are allowed, then spontaneous
pair-production of such remnants must occur in unbounded rate; and
(iii) Insights emerging from the AdS/CFT correspondence seem to support
the standard picture (A) of prompt information release via Hawking radiation.
\cite{AMPS}

How convincing are these points of objection? The claim (i) was actually
shown here to be invalid (provided that one accepts the semiclassicality
postulate). As for point (ii), the argument based on effective field
theory appears to be inconclusive (as briefly discussed in the previous
section). 

Point (iii) appears to be the most worrisome one. There appears to
be a deep conflict between the insights emerging from the AdS/CFT
correspondence, and the standard picture emerging from the equivalence
principle and the semiclassical theory of gravity. \cite{AMPS} 
We may hope that further investigations will clarify the roots of this conflict, hopefully retaining causality (at least at the macroscopic level) to the AdS/CFT
framework.

\subsection*{Acknowledgment}

I am grateful to Don Marolf for helpful discussions, and for clarifying
several issues in their recent paper.



\begin{thebibliography}{24}
\bibitem{PreskillA}Preskill \cite{Preskill} discussed Bank's arguments
\cite{horn} and raised several points of objection. Apparently none of them
is relevant to our case of 4D non-extremal BH, except possibly his
last point: Since semiclassical methods seems not to be sufficient
(when the remnant's throat becomes Planckian), {}``it is quite difficult
to calculate production rates ... or to otherwise support the contention
that the production rate is at odds with the effective field theory
viewpoint''. But this reservation actually reveals the non-conclusiveness
of the estimates of pair-production rates based on the effective field
theory.
\end{thebibliography}
\end{document}